\documentclass[conference]{IEEEtran}
\usepackage[utf8]{inputenc} %unicode support
\usepackage{graphicx}
\usepackage[square,sort,comma,numbers]{natbib}
\usepackage{booktabs, tabularx, multirow, array}
\newcommand{\otoprule}{\midrule[\heavyrulewidth]}
\usepackage{amsmath}
\usepackage{mathtools}
\usepackage{subcaption}
\usepackage[linesnumbered,ruled,vlined]{algorithm2e}
\usepackage{algpseudocode}
\usepackage{hyperref}
\usepackage[font={scriptsize}]{caption}
\usepackage{url}
\hypersetup{
    colorlinks,%
    citecolor=black,%
    filecolor=black,%
    linkcolor=black,%
    urlcolor=black
}

\ifCLASSINFOpdf
\else
\fi
\begin{document}
\title{\LARGE Fingerprinting Internet DNS Amplification DDoS Activities}
\author{\IEEEauthorblockN{Claude Fachkha, Elias Bou-Harb, Mourad Debbabi}
\IEEEauthorblockA{Computer Security Laboratory, CIISE, Concordia University \& NCFTA Canada \\
Montreal, QC, Canada\\
\{c\_fachkh, e\_bouh, debbabi\}@encs.concordia.ca}
}

%Telephone: (800) 555--1212, Fax: (888) 555--1212}
\maketitle
\begin{abstract}
%\boldmath
This work proposes a novel approach to infer and characterize Internet-scale DNS amplification DDoS attacks by leveraging the darknet space. Complementary to the pioneer work on inferring Distributed Denial of Service (DDoS) activities using darknet, this work shows that we can extract DDoS activities without relying on backscattered analysis. The aim of this work is to extract cyber security intelligence related to DNS Amplification DDoS activities such as detection period, attack duration, intensity, packet size, rate and geo-location in addition to various network-layer and flow-based insights. To achieve this task, the proposed approach exploits certain DDoS parameters to detect the attacks. We empirically evaluate the proposed approach using 720 GB of real darknet data collected from a /13 address space during a recent three months period. Our analysis reveals that the approach was successful in inferring significant DNS amplification DDoS activities including the recent prominent attack that targeted one of the largest anti-spam organizations. Moreover, the analysis disclosed the mechanism of such DNS amplification DDoS attacks. Further, the results uncover high-speed and stealthy attempts that were never previously documented. The case study of the largest DDoS attack in history lead to a better understanding of the nature and scale of this threat and can generate inferences that could contribute in detecting, preventing, assessing, mitigating and even attributing of DNS amplification DDoS activities.
\end{abstract}
% \fi
\IEEEpeerreviewmaketitle
\section{Introduction}
A DDoS attack is one of the major cyber attacks that attempts to make a computer or network
resources unavailable. DDoS activities, indeed, dominate today's attack landscape. In a recent report by Arbor Networks \cite{arbor}, it was concluded that 48\% of all cyber threats are DDoS. Governmental organizations, corporations as well as critical infrastructure were also recently deemed as DDoS victims \cite{govddos}. A DNS amplification attack is a form of DDoS that relies on the use of publically accessible open recursive DNS servers to overwhelm a victim system with DNS response traffic \cite{us-cert-dns-amplification}. A recent event demonstrated that even a cyber security organization became a victim of the largest (i.e., 300 Gbps) DNS amplification DDoS attack in history \cite{spamhaus-fact}. The above facts concur that DDoS attacks in general, and DNS amplification in particular, are and will continue to be a significant cyber security issue, causing momentous damage to a targeted victim as well as negatively affecting, by means of collateral damage, the network infrastructure (i.e., routers, links, etc.), the finance, the trust in, and the reputation of the organization under attack. In this work, we tackle the following questions: 1) How to infer large-scale DNS amplification DDoS activities? 2) what are the characteristics of DNS amplification DDoS attacks? and 3) what inferences can we extract from analyzing DNS amplification DDoS traces?

In this context, we frame this paper's contributions as follows:
\begin{itemize}
\item Proposing a systematic flow-based approach for inferring DNS amplification DDoS activities by leveraging DNS queries to darknets.
\item Characterizing the inferred DNS amplification DDoS threats during a recent 3 months period. 
\item Analyzing traces from the largest DNS amplification Attack of March 2013 \cite{spamhaus2} and uncovering the mechanism behind them.
\end{itemize}
The remainder of this paper is organized as follows:
In Section \ref{rel}, we survey the related work. In Section \ref{background}, we provide an overview and background information on DNS amplification attacks and darknet space. In Section \ref{methodology}, we present the proposed approach and elaborate on various aspects of its components. In Section \ref{EE}, we empirically evaluate the approach and disclose a case study on the largest DNS amplification DDoS attack. In Section \ref{lesson}, we discuss the lessons learned. Finally, Section \ref{conc} summarizes the paper and discusses the future work. 
\section{Related work}\label{rel}
First, the use of darknet to infer DDoS activities owes much to the pioneer work carried out by Moore et al. in \cite{Moore01inferringinternet} that was revisited in \cite{dosdetectmoore}. The key observation behind the authors' technique is that attackers, before executing a DDoS attack, spoof their addresses using random IPs. Hence, once the attack is executed, all the victims' replies (i.e., backscattered packets)  are bounced back to the fake IP addresses, which could be in the monitored darknet space. Their work is operated by CAIDA \cite{caida-ref}, which provide backscattered data for researchers. Numerous research work has been performed on such data to analyze DDoS activities. The majority focus on implementing new detection techniques to infer DDoS attacks \cite{fadlullah2010dtrab}, tracing-back the sources of attacks \cite{Yao:2010:PIT:2043164.1851237}, investigating spoofed attacks \cite{bi2010study} and visualizing attacks \cite{irwin2008high}. Our work is different from this category as their dataset is only based on reply packets and do not include request packets such as DNS queries. Hence, DNS amplified activities may not be inferred using their approach. 

Second, in the area of DNS traffic analysis, the most related work in this area is rendered by Oberheide et al. \cite{dark-dns-2007} who analyze DNS queries that target darknet sensors. The authors characterize these traces and propose a mechanism to implement a secure DNS service on darknet sensors. Moreover, Paxson \cite{Paxson:2001:AUR:505659.505664} is among the first to pinpoint the threats of DNS reflectors on making DDoS attacks harder to defend. In another work, Dagon et al. \cite{dagon2008corrupted} analyze corrupted DNS resolution paths and pinpoint an increase in malware that modified these paths and threatened DNS authorities. In comparison to our work, Oberheide et al. have not linked or investigated any DNS DDoS traces through their analysis but solely focused on analyzing DNS traffic. On the other hand, Paxson and Dagon et al. did not investigate darknet data. Therefore, all DNS amplification traces destined to unused IP addresses (darknet) cannot be detected through their analysis. However, darknet and other sources of data (i.e., Pasive DNS) could be associated to extract further intelligence on DNS amplification DDoS activities such as the approximate number of infections. Future work could consider the latter task.
\section{Background}\label{background}
In this section, we provide some background information related to the mechanism of DNS amplification attacks, the darknet space and DNS queries targeting the darknet.
\subsection{DNS Amplification DDoS Attacks}
%In a summary, DADDoS attack is a sophisticated type of DDoS attack. The attack cycle contains both request and reply packets. Requests are sent to the DNS servers whereas amplified replies are sent back to the victim. This attack can be executed in two stages. First, the attacker spoofs the IP address of the DNS resolver (e.g. host machine) and replaces it with the victim's address. Second, the attacker finds an Internet domain that is registered with many DNS records. During the attack, the attacker sends a large amount of spoofed DNS queries to the list of DNS records of that domain. This results in huge number of large size replies (each reply is up to 100 times greater than its corresponding request) to the victim. Attackers usually synchronize this process by controlling an army of zombies. Hence, this DDoS campaign amplifies even more the attack and bombards the victims with a denial of service.\\
A DNS amplification attack is a popular form of a DDoS, in which attackers use publically accessible open DNS servers to flood a target system with DNS response traffic \cite{us-cert-dns-amplification}. The primary technique consists of an attacker sending a DNS name lookup request to an open DNS server with the source address spoofed to be the target's address. When the DNS server sends the DNS record response, it is sent instead to the target. Attackers will typically submit a request for as much zone information as possible to maximize the amplification effect. In most attacks of this type, the spoofed queries sent by the attacker are of the type {\tt ANY}, which returns all known information about a DNS zone in a single request. Because the size of the response is considerably larger than the request, the attacker is able to increase the amount of traffic directed at the victim. By leveraging a botnet to produce a large number of spoofed DNS queries, an attacker can create an immense amount of traffic with little effort.
%\section*{Figures}
  %\begin{figure*}[h!]{images/scenario-dns.pdf}
	%\caption{DNS Amplification Scenario}
	%\label{dns-amplification-scenario}
	%\end{figure*}
%\subsection{Darknet Analysis}
%We answer the following questions to clarify some possible scenarios.
\subsection{Darknet Space}
%In our darknet analysis, the DADDoS detection is based on DNS queries destined to darknet sensor. The reason behind the assumption of finding only queries is that DADDoS amplified replies are aimed to be sent to a particular victim and not an unused dark IP. However, this assumption might be wrong in case of misconfiguration. If received by an open DNS resolver, all these DNS queries can trigger up to 100 times amplified replies to the requester. On the other hand, if the same queries are received to a secured DNS resolver (not open), then the replies will be dropped. This scenario is further depicted in Figure \ref{scenario1} and \ref{scenario2}.
In a nutshell, darknet traffic is Internet traffic destined to routable but unused Internet addresses (i.e., dark sensors). Since these addresses are unallocated, any traffic targeting such space is suspicious. Darknet analysis has shown to be an effective method to generate cyber threat intelligence \cite{dark1,6378947}. Darknet traffic is typically composed of three types of traffic, namely, scanning, backscattered and misconfiguration \cite{wustrowinternet}. Scanning arises from bots and worms while backscattered traffic commonly refers to unsolicited traffic that is the result of responses to DDoS attacks with spoofed source IP addresses. On the other hand, misconfiguration traffic is due to network/routing or hardware/software faults causing such traffic to be sent to the darknet sensors. 
\subsection{DNS Queries on Darknet}
On the darknet space, we observe a significant number of DNS queries that could be sent by the following sources:
\begin{itemize}
\item{Victim of Spoofed IP:} In this scenario, the attacker sends spoofed DNS queries on the Internet address space using the victim's IP address. All replies from the open DNS resolvers will bounce back towards the victim.
\item{Compromised Victim:} In this scenario, the attacker uses the victim's machine to send DNS queries. The attacker might use several techniques to control the victim's machine, including malware infection and/or vulnerability exploitation. This scenario do not involve spoofed DNS queries.  
\item{Scanner:} In this scenario, the attacker scans the Internet to infer the locations of open DNS resolvers. This task requires collecting information from the reply packets and hence, a non-spoofed address is used by the scanners.
 %Once a database of open Dis built, the attacker executes a DADDoS attacks, using a spoofed address or a compromised machine, by requesting only open DNS resolvers. 
\end{itemize}
In our work, we assert that high speed {\tt ANY} DNS queries will be sent from a victim of spoofed IP or/and compromised victim but not from a scanner. In other words, scanners might send {\tt ANY} DNS queries to the Internet but with low-speed rate to avoid receiving the amplified flood of replies.  
\section{Proposed Approach}\label{methodology}
This section presents and elaborates on our proposed approach that aims at inferring DNS amplification DDoS activities by leveraging darknet data. \emph{The approach exploits the idea of analyzing DNS queries that target the darknet space that were originally intended by the attacker to reach Internet open DNS resolvers}. The approach takes as input darknet traffic and outputs inferred DNS amplification DDoS insights. It is based on 2 components, namely, the detection and the rate classification components. We discuss these components in what follows.
\subsection{Detection Component}
The detection component takes as input darknet traffic and outputs DNS amplification DDoS flows. A flow is defined as a series of consecutive packets sharing the same source IP address targeting darknet addresses. To achieve the detection task, we base our detection component on analyzing DNS queries targeting darknet addresses. These DNS queries are attempts towards port 53. In order to detect DNS amplification DDoS, we built our approach in accordance with the parameters of Table \ref{parameters}.
\begin{table}[!h]
\centering
\begin{tabular}{p{2.5cm}p{3cm}}
\toprule%
%\multicolumn{6}{c}{\hspace{2cm} Prediction Techniques} \\\cmidrule{3-6}
\bf{Parameter} & \bf{Value} \\\otoprule  %\cmidrule{3-6}
%\multicolumn{2}{c} & Features &  &  & & \\\otoprule %
Packet Count & $>$ 25 \\\midrule
Scanned Hosts& $>$ 25 \\\midrule
%Average Rate & $>$ 0.5 (pps) \\\midrule
DNS Query Type & ANY \\\midrule
Requested Domain & Found in Root\_DNS\_DB  \\\midrule
\end{tabular}
\caption{DNS amplified DDoS Identification Parameters}
\label{parameters}
\end{table}
In this work, we build on top of \cite{dosdetectmoore} that inferred DDoS from darknet; we use 25 as a threshold for both the packet count and scanned hosts. 
This permits the filtering of misconfiguration traffic (i.e., a host sending many packets to only 1 unused IP address). Moreover, this verifies that the inferred DNS amplified attempts involve at least 25 distinct open DNS resolvers. Note that, we could have also added other parameters such as $attack$-$duration$ and $packet$-$rate$ to our detection component. However, we avoid using time-based constraints; we have detected some flash attempts \cite{Staniford:2004:TSF:1029618.1029624} that targeted thousands of distinct unused IPs within seconds and other stealthy scanning activities \cite{staniford2002practical} that persisted for several weeks.

In summary, our detection component labels a flow of traffic as a DNS amplification DDoS attack if it has sent at least 25 DNS query of type {\tt ANY} to distinct unused dark IP addresses. Further, the flow must have requested domains that exist in our root and TLD database.
\begin{figure*}[!t]
\centering
\includegraphics[width=0.85\textwidth]{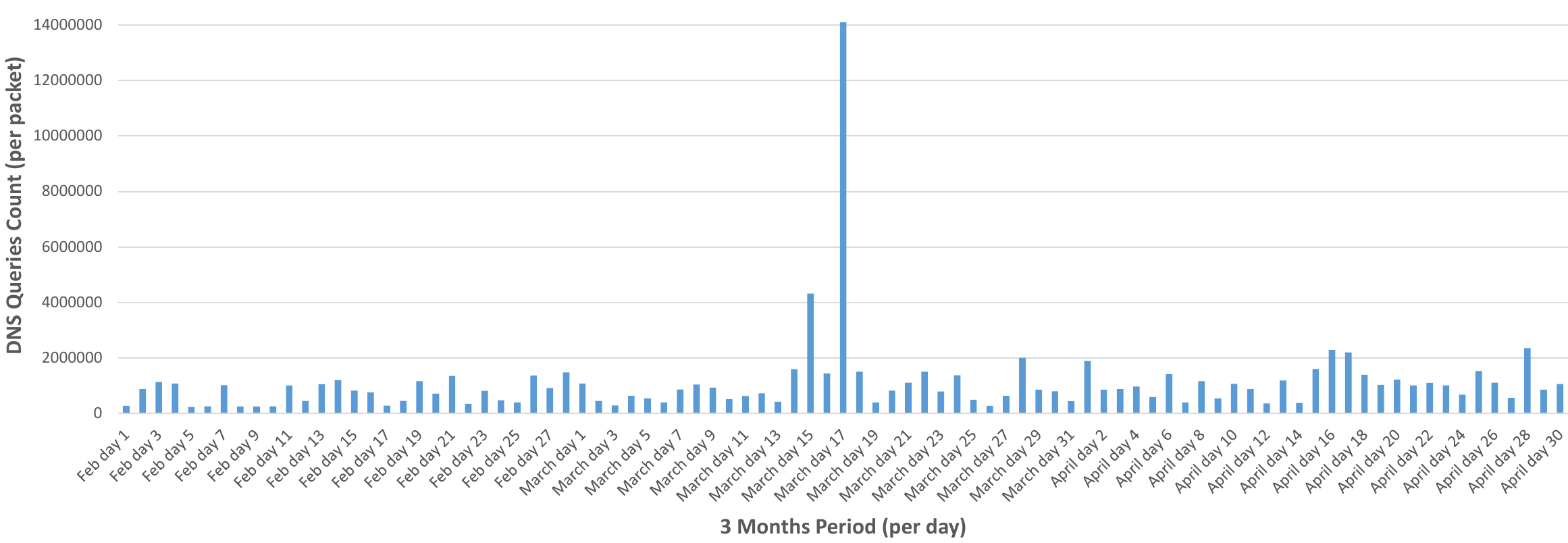}
\caption{DNS Queries Distribution of February, March and April 2013}
\label{tfs33}
\includegraphics[width=0.85\textwidth]{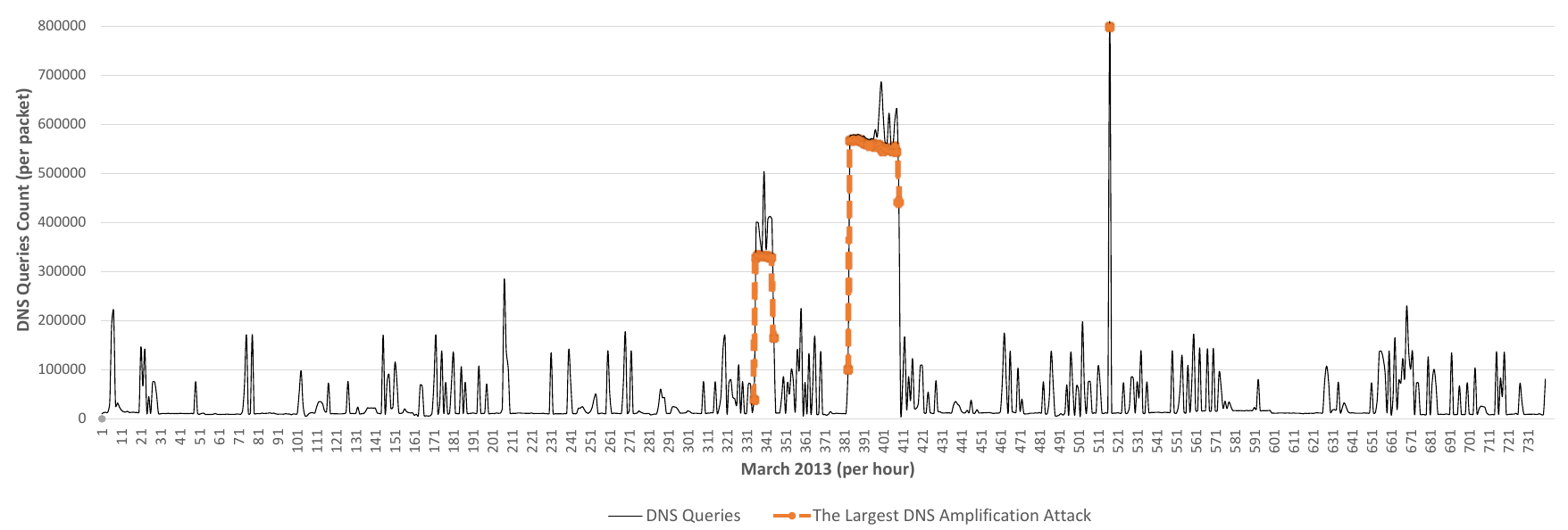}
\caption{DNS Queries Distribution of March 2013}
\label{tfs3}
\end{figure*}
\subsection{Rate Classification Component}
The rate of the attack is one of the major characteristics of DDoS activities \cite{dosdetectmoore}. After inferring DNS amplification flows, we noticed the existence of a large deviation among DNS amplification DDoS attack rates. For example, some flow rates reached more than 50 thousand packets per second (pps) whereas others were below 1 pps. Therefore, in order to understand more this large deviation and to group attacks per attack rates, we executed a rate classification exercise based on the values found in Table \ref{DADoS-classes}. \noindent
\begin{table}[!h]
\centering
\begin{tabular}{p{1.5cm}p{3cm}p{3cm}}
\toprule%
%\multicolumn{6}{c}{\hspace{2cm} Prediction Techniques} \\\cmidrule{3-6}
\bf{Attack Rate Category} & \bf{Value (pps)} \\\otoprule  %\cmidrule{3-6}
%\multicolumn{2}{c} & Features &  &  & & \\\otoprule %
Low & rate $\leq$ 0.5\\\midrule
Medium & 0.5 $<$ rate $<$ 4700\\\midrule
High & rate $\geq$ 4700\\\midrule
\end{tabular}
\caption{Attack Classification per Rate}
\label{DADoS-classes}
\end{table}
We have chosen a threshold of 0.5 pps to differentiate between low and medium attacks. This value is used in \cite{dosdetectmoore}. However, instead of neglecting such low rate attacks, similar to what \cite{dosdetectmoore} did, we adopt and exploit this value to detect a sub-category of attacks; stealthy attempts \cite{staniford2002practical}, which consists of slow scans that are generally hard to detect through flow-based detection and intrusion detection systems. In regards to high attack rate, this category contains high rate attempts that are commonly referred to as flash attacks \cite{Staniford:2004:TSF:1029618.1029624}. We have chosen a threshold of 4700 pps, which is the average rate of the Slammer worm propagation \cite{Staniford:2004:TSF:1029618.1029624}, to differentiate between medium and high rate attacks. In this exercise, we assume that the average rate of the fastest worm propagation in 2003 will have, at least, similar rates as flash attacks in 2013. Please note that in general, worm propagation performs scans for vulnerabilities on hosts in an attempt to exploit or infect the victims. In comparison to the DNS amplification DDoS attempts, the attackers generate a similar portsweep propagation attempts to detect open DNS resolvers and execute the attack in one shot. The latter technique does not aim at searching for a vulnerability to exploit, but instead sends benign DNS {\tt ANY} queries to abuse the open DNS resolver services and amplify the reply on a victim. 
\section{Empirical Evaluation}\label{EE}
The evaluation is based on a real darknet dataset during a 3 months period, namely, February, March and April, 2013. In general, we possess real darknet data that we receive on a daily basis from a trusted third party. The darknet sensors monitor /13 address blocks (i.e., $\approx$ half a million dark IPs). The analyzed data consists of an average of 720 GB of one-way communications to unused IPs. In regards to our characterization tasks, we used several network-based monitoring and statistical tools such as TCPdump. We abide and closely follow the steps of our proposed approach that was discussed in Section \ref{methodology} to elaborate on our analysis, which is based on three main elements, namely, the characterization and a case study. In total, our approach identified a total of 134 DNS amplification DDoS attacks including high-speed, medium and stealthy attacks.
\subsection{DNS Amplification DDoS Characterization}\label{profiling}
In this section, we present the overall DNS amplification DDoS statistics related to our analyzed dataset. The overall DNS queries distribution is shown in Figure \ref{tfs33}. The outcome clearly fingerprints the largest DNS amplified DDoS attack that occurred in March 2013 \cite{spamhaus2}. On the other hand, in order to have a closer look at this attack, we depict Figure \ref{tfs3} that illustrates the distribution of the queries for the month of March. The average DNS queries arrival time per hour is approximately 58050 packets. Obviously, several large-scale DNS Amplified DDoS attacks caused some peaks at some periods such as at hours 340, 400 and 517 in which the distribution of packets was raised to 503995, 686774 and 798192 packets, respectively. More explanation on these peaks are discussed in Section \ref{insights-casestudies}.
\subsubsection{Query Type Distribution}
In order to understand the types of DNS queries received on our dark space, we list in Table \ref{summary-dns-query-type} the DNS query type distribution of the analyzed dataset. As expected, the vast majority of these are {\tt ANY} queries. Note that the top 4 records are the same for the entire 3 months period. Further, in contrast with the results in 2007 by \cite{dark-dns-2007}, that found that {\tt ANY} records scored only 0.0199\% of the entire perceived records, we record 52.23\% as observed on the darknet space. As a result, we can safely assume that the recent trend of DNS amplification attacks are behind the increase of {\tt ANY} records found on the darknet in the current year \cite{spamhaus2}.
\begin{table}[!h]
\centering
\begin{tabular}{p{1.8cm}p{1.6cm}p{1.6cm}}
\toprule%
%\multicolumn{6}{c}{\hspace{2cm} Prediction Techniques} \\\cmidrule{3-6}
\bf{February Packet \newline Count (\%)} & \bf{March Packet Count (\%)} & \bf{April Packet Count (\%)} \\\otoprule  %\cmidrule{3-6}
%\multicolumn{2}{c} & Features &  &  & & \\\otoprule %
10047038 \newline A (49.02\%) & 27649274 \newline ANY (64.23\%)& 18378685 \newline ANY (54.60\%) \\\midrule
7763817 \newline ANY (37.88\%)& 11310058 \newline A (26.28\%)&  11595908\newline  A (34.45\%)\\\midrule
2479572 \newline TXT (12.10\%)& 2459257  \newline TXT (5.71\%)& 3402073 \newline TXT	(10.11\%) \\\midrule
100463 \newline MX (0.49\%)& 500143 	\newline  MX 	(1.16\%)&  180779\newline  MX (0.54\%)\\\midrule
29232 \newline PTR (0.14\%)& 63340 	\newline  RRSIG 	(0.15\%)&  28716 \newline AAAA (0.09\%)\\\midrule
\end{tabular}
\caption{Top 5 DNS Query Type Distribution of 3 Months Period}
\label{summary-dns-query-type}
\end{table}
\subsubsection{Requested Domains}
In our analysis, we found that Root is the most requested domain name as perceived by the monitored darknet. Recall that attackers will typically submit a request for as much zone information as possible to maximize the amplification effect. Note that, from our data, the second top requested domain belongs to one of the largest Internet-scale DNS operators. 
\begin{table*}[!t]
\centering
\resizebox{120mm}{!}{
\begin{tabular}{p{1.5cm}p{1.5cm}p{1.7cm}p{1.7cm}p{1cm}p{1.5cm}p{1cm}p{1cm}p{1cm}}
\toprule%
%\multicolumn{6}{c}{\hspace{2cm} Prediction Techniques} \\\cmidrule{3-6}
\bf{Victim} & \bf{Requested Domain Name} & \bf{Detection Period} & \bf{Analyzed Attack \newline Duration (second)} & \bf{Intensity (packet)} &  \bf{Contacted Unique \newline Dark IPs} & \bf{Avg. Packet Size (Bytes)}& \bf{Avg. Rate (pps)}  &\bf{Rate \newline Category}\\\otoprule  %\cmidrule{3-6}
M1 &B & March 15 & 34605 & 3176785 & 360683  & 68.00 & 91.80  & Medium\\\midrule
M2 & B & March 17 to 18 & 93508 & 14464427 & 360705 & 68.00& 154.69 & Medium\\\midrule
\end{tabular}}
\caption{DNS Amplification DDoS Traces}
\label{summary-low}
\end{table*}
\subsection{Case Study}\label{insights-casestudies}
Out of the 134 detected attacks, we discuss in this section one of the major case studies that belong to a medium speed attack. The latter is one of the major inferred DNS amplification DDoS in terms of size and impact. This attack targeted one victim using 2 hosts (ID M1 and M2 of Table \ref{summary-low}). This attack scanned around 360000 unique dark IPs (68\% of the monitored /13 darknet), and hence could be considered the most comprehensive compared to all other threats. Our analysis linked these traces to the largest DNS amplification DDoS \cite{spamhaus2} for the following reasons: 1) in addition to the use of the {\tt ANY} DNS query , the traces of this attack targeted the "`ripe.net"' domain name; this domain was used in the largest DDoS as declared in a blog posted by the victim \cite{spamhaus2}; 2) the timing of the traces from the host with ID 1 started on March 15$^{\mbox {th}}$, whereas those of the host with ID 2 started on March 17$^{\mbox {th}}$. The two mentioned dates could be found in the media \cite{bloomberg,irish} and were posted on Twitter on March 17$^{\mbox {th}}$ by a company support personnel \cite{twitter}. In order to depict this distributed attack, in Figure \ref{tfs3}, we highlighted the threat using a colored dashed-line. The first or/and second peaks are likely performed as testing before actually executing the largest DDoS as demonstrated by the third peak. Our result match the ascending order of peaks as discussed by the victims \cite{spamhaus2}. This case study is probably sent by an attacker using spoofed IP address of the victims or using compromised machines; we unlikely consider these activities as scanning events that are using legitimate addresses (i.e., the intention is not to DDoS themselves but other targeted victims).

In addition to performing several validation of our results through DShield and the media, we execute a renowned Network Intrusion and Detection System (NIDS) (i.e., Snort) on the whole traces to see if we can detect such malicious activities. The NIDS labeled 129 out of the inferred 134 (96\%) DNS amplification DDoS as executing filtered portsweep probes. We have found that the 5 undetected attacks refer to the low-speed (stealthy) attacks, which are, by default, undetectable using a typical NIDS. In summary, we can claim that our approach that aims at inferring DNS amplification DDoS yielded zero false negative in comparison with a leading NIDS. Further, our approach, leveraging the darknet space, can infer DNS amplified DDoS activities while a NIDS is limited to pinpointing scanning attempts.

\section{Lessons Learned}\label{lesson}
From this work, we can extract the following insights related to DNS amplification attacks: First, when compared to previous years, we have found that the DNS amplification attacks are behind the increase of DNS queries of type {\tt ANY} on the Internet. Second, we have pinpointed that the majority of the attacks target the root domain. Third, we have encountered that DNS amplified attack rates can range from very low to high speeds. High speeds attacks pinpoint victims of spoofed attacks and compromised machines whereas the very slow attacks reflects stealthy scans. Last but not least, we have unexpectedly uncover a UDP-based mechanism used by DNS amplification attackers to execute DNS amplification attacks in a highly rapid manner without collecting information about open DNS resolvers. Further, more importantly, we have inferred that unlike typical DDoS attempts that scan for vulnerable machines and then execute the attack, the largest DNS amplification analyzed was executed in only one step;  DNS queries are sent to the Internet with the intention to reach open DNS resolvers, which subsequently trigger an amplified reply to the victim.


\section{Conclusion}\label{conc}
This work presented a new approach to infer Internet DNS Amplification Denial of Service activities by leveraging the darknet space. The approach corroborated the fact that one can infer DDoS attacks without relying on backscattered analysis. The detection module is based on certain parameters to fingerprint network flows as DNS amplification DDoS related. The classification module amalgamates the attacks based on their possessed rate. The analysis was based on 720 GB of real darknet traffic collected during a recent 3 months period. The results disclose 134 DNS amplified DDoS activities, including flash and stealthy attacks. Moreover, the case study provided significant cyber security intelligence related to the largest DNS amplification attack. As for future work, we aim to execute our model on a larger data set and implement our proposed approach in a near real-time fashion.


%\bibliographystyle{IEEEtran}
\bibliographystyle{IEEEtran}
\bibliography{references}

\end{document}